\documentclass[
reprint,
prd,
floatfix,
]{revtex4-2}

\usepackage{graphicx}% Include figure files
\usepackage{xcolor}

\newcommand{\araa}{Annual Review of Astronomy and Astrophysics}

\newcommand{\apjl}{Astrophys. J. Lett.}
\newcommand{\aap}{Astron. Astrophys.}

\newcommand{\mnras}{Mon. Notices Royal Astron. Soc.}
\newcommand{\jcap}{J.~Cosmology Astropart. Phys.}

\newcommand{\be}{\begin{equation}}
\newcommand{\ee}{\end{equation}}
\def\ba{\begin{eqnarray}}
\def\ea{\end{eqnarray}}

\begin{document}

\title{Lensing by primordial black holes: constraints from gravitational wave observations}

\author{Jie-Shuang Wang}
\affiliation{Tsung-Dao Lee Institute, Shanghai Jiao Tong University, Shanghai 200240, China; jiesh.wang@gmail.com }
\author{Antonio Herrera-Mart\'in} 
\affiliation{School of Physical and Chemical Sciences, University of Canterbury, Private Bag 4800, Christchurch, New Zealand; antonio.herreramartin@canterbury.ac.nz}
\author{Yi-Ming Hu} 
\affiliation{TianQin Research Center for Gravitational Physics and School of Physics and Astronomy, Sun Yat-sen University (Zhuhai Campus), 2 Daxue Road, Zhuhai 519082, China; huyiming@mail.sysu.edu.cn}

%\shorttitle{Lensing of GW by PBHs}
%\shortauthors{Wang et al.}

\begin{abstract}
Primordial black holes (PBHs) have been proposed to explain at least a portion of dark matter. Observations have put strong constraints on PBHs in terms of the fraction of dark matter which they can represent, $f_{\rm PBH}$, across a wide mass range -- apart from the stellar-mass range of $20M_\odot\lesssim M_{\rm PBH}\lesssim 100M_\odot$. In this paper, we explore the possibility that such PBHs could serve as point-mass lenses capable of altering the gravitational-wave (GW) signals observed from binary black hole (BBH) mergers along their line-of-sight. We find that careful GW data analysis could verify the existence of such PBHs based on the {\em fitting factor\/} and odds ratio analyses. 
When such a lensed GW signal is detected, we expect to be able to measure the redshifted mass of the lens with a relative error $\Delta M_{\rm PBH}/M_{\rm PBH}\lesssim0.3$.
If no such lensed GW events were detected despite the operation of sensitive GW detectors accumulating large numbers of BBH mergers, it would translate into a stringent constraint of $f_{\rm PBH}\lesssim 10^{-2}-10^{-5}$ for PBHs with a mass larger than $\sim10M_\odot$ by the Einstein Telescope after one year of running, and $f_{\rm PBH}\lesssim 0.2$ for PBHs with mass greater than $\sim 50M_\odot$ for advanced LIGO after ten years of running. 
\end{abstract}

\maketitle
%%%%%%%%%%%%%%%%%%%%%%%%%%%%%%%%%%%%%%%%%%%%%
%%%%%%%%%%%%%%%%%%%%%%%%%%%%%%%%%%%%%%%%%%%%%
\section{Introduction}\label{sec:intro}
%Recent study found that 
Dark matter accounts for around a quarter of the total energy density based on the standard cosmological model \citep{Planck2016}. 
However, its nature is still vastly unknown. One possible explanation for this is that dark matter consists of primordial black holes (PBHs) formed in the early Universe \citep[e.g.][]{Zel1967,Hawking1971,Carr1974,Meszaros1974,Chapline1975,Cai2018,Ando2018}, which can act as massive astrophysical compact halo objects. 

The fraction of PBHs to the total amount of dark matter, $f_{\rm PBH}$, whose mass range crosses $\sim33$ orders of magnitude, has been constrained by various observations (e.g. see \citet{Carr2016} and references therein). More specifically, strong limits on the abundance of PBHs have been placed in the mass range $10^{-6}-20\,M_\odot$ due to the non-detection of microlensing events from nearby galaxies \citep{Alcock2001,Tisserand2007}, 
and in the range $>100\,M_\odot$ from the observation of wide halo binaries and ultra-faint dwarf galaxies \citep{Quinn2009,Brandt2016,Li2017}, and the cosmic microwave background \citep{Ali2017,Bernal2017}. 

However, only weak constraints have been placed in the window $20M_\odot\lesssim M_{\rm PBH}\lesssim 100M_\odot$.
Intriguingly, this mass range is consistent with that of the black holes (BHs) recently detected by the LIGO and Virgo observatories \citep{Abbott2016a,Abbott2016b,Abbott2016c,Abbott2017a,Abbott2017b,Abbott2017c,Ligo2018a}.
Therefore, it has been proposed that PBHs, as cold dark matter, could form binaries to account for the observed gravitational wave (GW) events \citep{Bird2016}. 
Moreover, great progress has been made to probe the PBHs in this mass range \citep{Munoz2016,Kovetz2017,Ji2018,Jung2019}. 
For example, it is suggested that the lensing effect on the light curves of fast radio bursts (FRBs) \citep{Munoz2016} and gamma-ray bursts (GRBs) \citep{Ji2018} can constrain $f_{\rm PBH}$. Yet it should be noted that the distances of FRBs and GRBs are usually unknown. 

It is also suggested that lensing of GWs can be used to study PBHs \citep[e.g.][]{Jung2019,Liao2020,Diego2020}. 
Lensing of GW has been studied for different situations in terms of lens properties \citep[e.g.][]{Nakamura1998,Nakamura1999,Takahashi2003,Cao2014,Fan2017PhRvL.118i1102F,Liao2017NatCo...8.1148L,Dai2018,Christian2018,Lai2018,Oguri2018,Jung2019,Diego2019,Hou2020,Diego2020,Buscicchio2020PhRvL.125n1102B,Ezquiaga2021PhRvD.103f4047E,Cheung2021,Xu2021}.
Yet up to now, no compelling evidence for lensing has been found in the observed GW signals \citep{Hannuksela2019,LVC2021lensing}.
For lenses of Schwarzschild radii smaller than or comparable to the GW wavelength, diffraction \citep[e.g.][]{Nakamura1998,Nakamura1999,Dai2018} and interference \citep[e.g.][]{Takahashi2003,Cao2014,Christian2018} effects can be important, although the magnification may be insignificant. 
In the case of ground-based detectors with frequencies $f\sim10-10^3$~Hz, lenses with mass $\lesssim 10^2 - 10^5M_\odot$ can lead to such wave effects. 
If a fraction of dark matter is made up of PBHs with a mass of $\sim20-100M_\odot$, microlensing events by diffusely distributed PBHs would be expected. 
There are also cases where the macrolensing effect induced by the cosmic structures hosting those PBHs, such as galaxies and clusters, may be of importance. And such macrolensing effect can be further influenced by the microlensing effect by individual PBHs at sufficiently large magnification \citep{Diego2019,Diego2020,Cheung2021}.

In this paper, we explore the possibility of probing stellar-mass PBHs through their strong gravitational lensing effect on GW events produced by the mergers of binary black holes (BBHs). 
We assume the PBHs are uniformly distributed in the Universe and focus on their microlensing effect. 
To quantify the difference between the lensed and unlensed waveforms, the residual of SNR is used as an indicator in previous research~\citep[e.g.][]{Jung2019}.
One can, in principle, adopt the full Bayesian model selection method to distinguish the lensed signal from the unlensed \citep[e.g.][]{Lai2018,Christian2018,Hannuksela2019,LVC2021lensing}, which would be too time-consuming for a rate study.
Therefore, we adopt a criterion based on the {\em fitting~factor} \citep[e.g.][]{Diego2019,Cheung2021}, which is suitable for such rate study. But we validate our choice through a comparison between the Bayesian model selection and the {\em fitting~factor}. As the lens model involves more parameters, a model selection criterion such as the odds ratio, which penalizes more complicated models, is a more desirable approach. By validating through the odds ratio calculation, we can perform a fast calculation on top of the {\em fitting~factor} and obtain conservative constraints.

The third-generation GW detectors, such as Einstein Telescope (ET) \citep{Abernathy2011} and Cosmic Explorer \citep{Abbott2017d}, could detect GW events at large distances, underlining their great potential for placing stringent constraints on $f_{\rm PBH}$.
Furthermore, compared with the constraining effect of microlensing on FRBs and GRBs, the advantage of GW events is that the luminosity distance of GW events can be obtained from GW data. Consequently, the lensing of GW events can constrain the PBH fraction more straightforwardly and effectively. 
In section \ref{lensingeffect}, we study the lensing effect on a GW signal, which involves the wave effects. In Section \ref{FF_ODDS}, we show how to distinguish the lensed GW signal from the unlensed one. 
We study the lensing probability in Section \ref{probablity} and present the constraint on $f_{\rm PBH}$ in Section \ref{constraint}. Our conclusion and discussion are shown in Section \ref{conclusion}.

\section{The effect of gravitational lensing on a GW signal}\label{lensingeffect}

The inspiral-merger-ringdown GW waveform from a BBH merger event can be described by a phenomenological GW waveform model \citep{Ajith2011}. Considering that the BBH system consists of two BHs with redshifted masses $m_1(1+z_{\rm S})$ and $m_2(1+z_{\rm S})$, the waveform in the frequency domain can be described as $h(f,\theta_p,\theta_a)$, where $f$ is the GW frequency, $\theta_p$ comprises the intrinsic parameters that describe the phase -- including the redshift of the BBH system $z_{\rm S}$, the redshifted total mass $M_{\rm BBH,z}=M_{\rm BBH}(1+z_{\rm S})=(m_1+m_2)(1+z_{\rm S})$, symmetric mass ratio $\eta = m_1m_2/(m_1+m_2)^2$,
dimensionless effective spin parameter $\chi$, and the nominal coalescence time and phase $t_0,~\phi_0$, which serve as two integration constants (see Eq. 1 in \citet{Ajith2011}).
Note the effect of spin misalignment is not included in this waveform model \footnote{It is suggested that the spin misalignment may bring difficulties to identify the microlensing effect, but its quantitative effect remains to be studied.}. 
And $\theta_a$ includes the extrinsic parameters that describe the amplitudes only, i.e. the distance of the source in the observation frame $D$, the sky position ($\theta,~\phi$), the polarization angle ($\psi$), and the inclination angle ($\iota$) of the binary system. Throughout this paper, the waveform models are calculated by averaging over all these four angles ($\theta,~\phi,~\psi,~\iota$).

Up-to-now, more than forty BBH merger events have been officially reported by LIGO and Virgo Collaboration \citep{Abbott2016a,Abbott2016b,Abbott2016c,Abbott2017a,Abbott2017b,Abbott2017c,Ligo2018a,GW190412,GW190521,Abbott20202nd}. 
Throughout this paper, we adopt averaged representative values of $\eta=0.238$ and $\chi=0.024$ from the GWTC-1 catalog that presents GW events detected during the first and second observing runs \citep{Ligo2018a}.  These values are also roughly consistent with results from the second LIGO Virgo catalog, GWTC-2 \citep{Abbott20202nd}.
The specific choice of these parameters does not change our conclusions significantly but does greatly simplify our analysis.
With the symmetric mass ratio and the spin parameter fixed, the signal-to-noise ratio (SNR, denoted by $\rho$) depends only on the total mass and the redshift. 
A horizon redshift ($z_{\rm S, c}$) is then defined at which the GW event with critical mass $M_{\rm BBH, c}$ will have a SNR $\rho(M_{\rm BBH, c},z_{\rm S, c})=8$.

\begin{figure}
	\centering
	\includegraphics[width=0.5\textwidth]{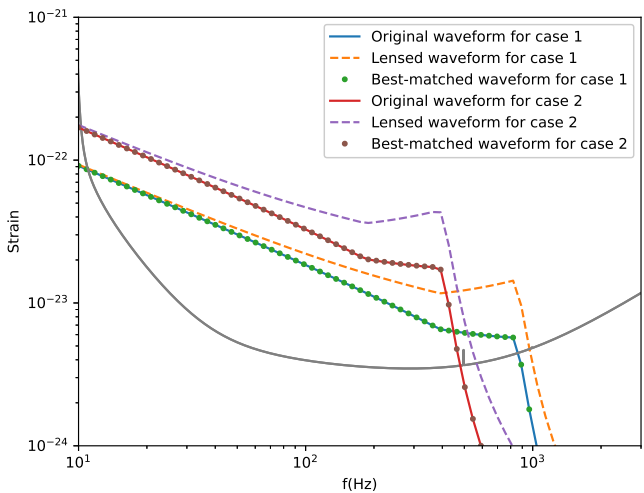}
	\caption{The original, lensed, and best-matched (using $FF$ analyses) waveforms for two cases with BBH mass $M_{\rm BBH,z}=20M_{\odot}~(\rm{case~ 1}),~42.5M_{\odot}~(\rm{case~2})$.
	We fixed $\eta=0.238$, $\chi=0.024$, $M_{\rm L,z}=42.5M_{\odot}$, and $y=0.1$.
	}\label{fig_lens}
\end{figure}

We focus on the microlensing effect due to stellar-mass PBHs, where their Schwarzschild radii are comparable to the GW wavelength. We adopt a wave-optics treatment to calculate the modification due to lensing \citep{Peters1974,Schneider1992,Takahashi2003,Cao2014}.
When the gravitational wave signal propagates under the gravitational field of a point-mass lens with mass $M_{\rm L}$ at redshift $z_{\rm L}$, the {modified} waveform is 
\be
h_l(f,\theta_l,\theta_p,\theta_a)=F(f,\theta_l)h(f,\theta_p,\theta_a),
\ee where the amplification function is \citep{Peters1974,Takahashi2003},
\ba
  F(f,\theta_l) = \exp\left[ {\pi w\over 4} + i{w\over 2}
 \left( \ln \left(  {w\over 2} \right)
 -2 \phi_m(y) \right)\right] \nonumber\\
 \times \Gamma \left( 1- \frac{i}{2} w \right) ~_1 F_1 \left(
 {i\over 2} w,1; {i\over 2} wy^2 \right), \label{eq:F_f}
\ea
and $\theta_l$ includes the lensing parameters $(M_{\rm L,z},~y)$, 
$w=8 \pi M_{\rm L,z} f$, $M_{\rm L,z}=M_{\rm L} (1+z_{\rm L})$ is the redshifted mass of the point-mass lens, $\phi_m(y)=(x_m-y)^2/2-\ln{x_m}$, $x_m=(y+\sqrt{y^2+4})/2$, $y=\beta/\beta_E$ is the angular impact parameter normalized by the angular Einstein radius $\beta_E$, and $_1 F_1$ is the confluent hyper-geometric function. 
We show two examples in Fig.~\ref{fig_lens}, where the original (unlensed) waveforms are obtained by setting $M_{\rm BBH,z}=20M_{\odot}~(\rm{case~ 1}),~42.5M_{\odot}~(\rm{case~2})$, $\eta=0.238$, $\chi=0.024$. While the lensed waveforms are calculated with parameters $M_{\rm L,z}=42.5M_{\odot}$ and $y=0.1$. It indicates that the lensing effect is more significantly at higher frequencies. 

\section{Distinguishing the lensing effect within GW data}\label{FF_ODDS}

Once a GW event is detected, one can search for the possible microlensing effect by replacing the unlensed templates with the lensed ones.
Suppose that there is a lensed GW signal with a waveform $h_l=F(f,\theta_l)h(f,\theta_p,\theta_a)$ received by the detectors while both the unlensed waveform templates ($h'(f,\theta_p',\theta_a')$) and lensed templates ($h_l$) are used in the data analyses. 
We adopt the {\em Fitting Factor\/} (denoted by $FF$, with $FF\leq 1$) and the odds ratio
to indicate the capability of identifying the microlensing effect following previous research \citep[e.g.][]{Takahashi2003,Cao2014,Lai2018,Christian2018,Cheung2021}. The $FF$ is initially used to describe the match between different waveform models, or between waveform models and true signals \citep[e.g.][]{Sathyaprakash1991,Poisson1995,Owen1996,Damour1998,Ajith2011,Babak2013,Harry2014,Canton2014}. 
Here it is defined as the maximum match between the lensed GW waveform and the unlensed waveform ($h'(\theta_p',\theta_a')$) over the template bank \citep[e.g.][]{Takahashi2003,Cao2014,Lai2018,Cheung2021}, 
\be
FF=\mathop{\text{max}}_{\theta_p',\theta_a'}{\Big(h'(\theta_p',\theta_a')\Big|h_l\Big)\over\sqrt{(h'|h')(h_l|h_l)}},\label{FF_analysis}
\ee
where the noise-weighted inner product of the waveform is
\be
(h_1|h_2)=4\text{Re}\int_{0}^{\infty}h_1(f)h_2^*(f){{\rm d}f\over S(f)},
\ee 
%$^*$ represents the complex conjugate, 
and $S(f)$ is the power spectrum density (PSD) of the detector. 
In this work we perform two case studies by adopting the PSDs of the advanced LIGO (aLIGO) \citep{Dwyer2015} and of the ET-D \citep{Abernathy2011,Abbott2017d}.

\begin{figure}
	\centering
	\includegraphics[width=0.49\textwidth]{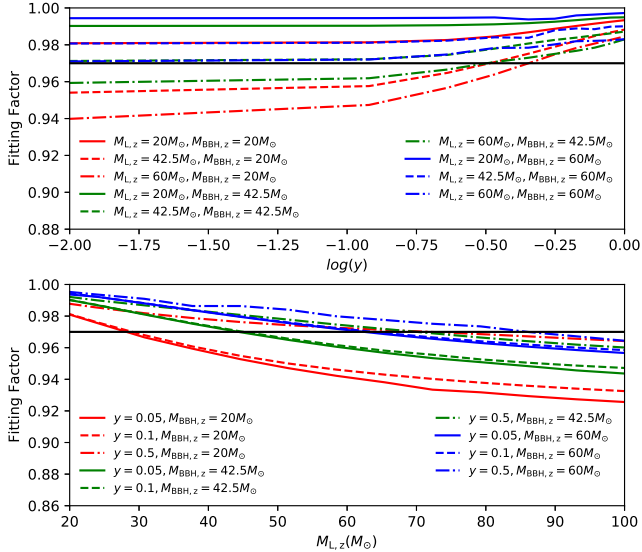}
	\caption{The fitting factor ($FF$) as a function $y$ (top panel) or $M_{\rm L}$ 
		(bottom panel) based on the PSD of aLIGO. We fixed $\eta=0.238$ and $\chi=0.024$. While the BBH mass is chosen to be $M_{\rm BBH,z}=20,~42.5,~60~M_{\odot}$. 
		Three different cases with lensing object mass 
		$M_{\rm L,z}=20,~42.5,~60~M_{\odot}$ are shown in top
		panel, and three different cases with impact parameter $y=0.05,~0.1,~0.5$ are 
		presented in bottom panel. The black line is the benchmark value of $FF=0.97$.
	}\label{fig_FF}
\end{figure}

To calculate the $FF$, we simultaneously vary the parameters $t_0,~\phi_0,~M_{\rm BBH,z},~\eta,~\chi$ of the unlensed waveform ($h'$) over the template bank to obtain the best-matched waveform. 
As discussed in Section \ref{lensingeffect}, we fixed the parameters $\eta$ and $\chi$ to their average values based on the GWTC-1 and GWTC-2 catalogs, so that the lensed waveform and the $FF$ are mainly determined by the lensing parameters and the redshifted BBH mass, i.e. $h_l=F(f,M_{\rm L,z},y)h(f,M_{\rm BBH,z})$, and $FF=FF(M_{\rm BBH,z},M_{\rm L,z},y)$. 
Two exemplary cases of $FF$ analyses are presented in Fig. \ref{fig_lens}, which shows that there is only a little difference between the best-matched waveforms obtained through $FF$ analyses (Eq. \ref{FF_analysis}) and the original unlensed waveforms, consistent with previous research \citep{Cao2014,Dai2017,Smith2018}.
We then study the evolution of $FF$ over large parameter space.
As examples, we show the dependence of $FF$ with $y$ and $M_{\rm L,z}$ for several cases of the redshifted mass $M_{\rm BBH,z}=20,~42.5,~60~M_{\odot}$ in Fig. \ref{fig_FF}. 
The BBH parameters act on $FF$ mainly through the maximum frequency of the BBH merger source $f_{\rm max}\propto M_{\rm BBH,z}^{-1}$ \citep{Ajith2011}. 
The PBH lens distorts the waveform more strongly at higher frequencies, so that the $FF$ correlates with $M_{\rm BBH,z}$ positively. 
Also for higher redshifted lens mass $M_{\rm L,z}$ or smaller impact parameter $y$, the waveform tends to be more significantly distorted, so that the $FF$ has a negative correlation with $M_{\rm L,z}$, and a positive correlation with $y$ as shown in Fig. \ref{fig_FF}. 
Note however for $y\lesssim 0.1$, the $FF$ changes very slowly, and is almost invariant.

The threshold at which two waveform models can be distinguished is usually set to $FF=0.97$; this corresponds to a loss of less than 10\% of signals \citep{Owen1996,Damour1998,Ajith2011,Babak2013,Harry2014,Canton2014,Lai2018}. 
This threshold is introduced to indicate the effectiveness of template models of GWs from BBH mergers. However, calculations of the Akaike Information Criteria (AIC;~\citep{Akaike1974}) indicates that this threshold will generally lead to a $|\Delta {\rm AIC}|>6.9$ between the lensed and unlensed waveforms as shown in the Appendix \ref{appendix:FF}. And the odds ratio analysis, as will be shown below, also support the validity of this threshold. 
Therefore, we also adopt this threshold.
The available parameter space for distinguishable lensing signals is then obtained by numerically searching for the critical parameters that makes
\begin{equation}
    FF(M_{\rm BBH,z},M_{\rm L,z},~y_c)= 0.97,
\end{equation}
in the parameter space of $M_{\rm BBH,z},M_{\rm L,z}> 10M_\odot$ and $y_c\in [0.01,2]$. 
We find the threshold can be met in all cases for the impact parameter and the redshifted lens mass satisfying
\begin{equation}
    y_c\approx 0.1-0.3~\rm{and} ~M_{\rm L,z} \approx (1.0-1.6) M_{\rm BBH,z}. \label{eq:FFthreshold}
\end{equation}
As examples, the cases for GW events with mass $M_{\rm BBH,z}=20,~42.5,~60~M_{\odot}$ are shown in Fig. \ref{fig_FF}.

\begin{figure}
	\centering
	\includegraphics[width=0.49\textwidth]{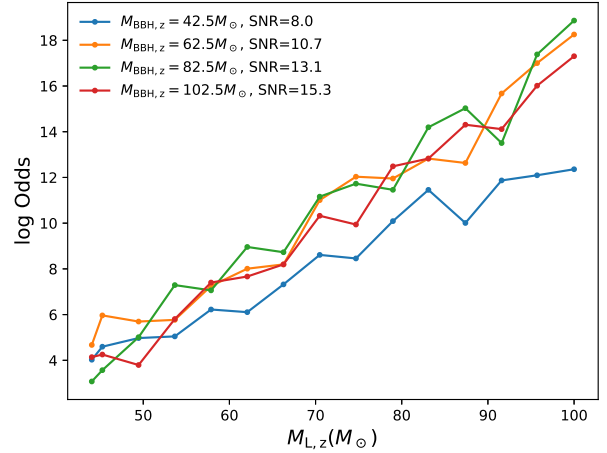}
	\caption{
	We fixed $\eta=0.238$, $\chi=0.024$, and the BBH mass is chosen to be $M_{\rm BBH,z}=42.5,~62.5,~82.5,~102.5~M_{\odot}$. 
	These sources are assumed at the same redshift ($z_{\rm S, c}$) so that the SNRs of the unlensed signals are ${\rm SNR}=8.0,~10.7,~13.1,~15.3$. 
	The Odds ratios are calculated for different $M_{\rm L,z}$ based on the PSD of aLIGO and $y=0.1$. 
	}\label{fig_odds}
\end{figure}

The above result shows that GW events of mass $M_{\rm BBH, c}$ within their horizon redshift ($z_{\rm S, c}$) can be used to probe PBHs of redshifted mass $M_{\rm L,z}\gtrsim M_{\rm BBH, c}(1+z_{\rm S, c})$. 
While for GW events of mass $M_{\rm BBH}>M_{\rm BBH, c}$ at $z_{\rm S, c}$ will have higher SNR than the events of $M_{\rm BBH}=M_{\rm BBH, c}$. The effect of the SNR on the distinguishability is not included in the $FF$ analysis, while it has been found that it can help to amplify and differentiate the lensing effect. 
Thus, following Ref. \citep{Christian2018}, we adopt the odds ratio analysis to include this SNR effect,
where we use $\ln\mathcal{L} \approx-\rho^2(1-FF)$ to construct the likelihood~\citep{Lange2017}. Maximizing the likelihood reduces the mismatch of the waveforms, and the best match is consistent with~\citep{Lindblom2008}. 
We set the minimum requirement to recognize the lensed signal to be \begin{equation}
    \log \rm{Odds}(M_{\rm BBH,z},M_{\rm L,z},~y_c) = 4.
\end{equation}
For a given event of $M_{\rm BBH,z}\geq M_{\rm BBH, c}(1+z_{\rm S, c})$, we solve the above equation numerically by calculating the probability for the best combination of parameters for the lensed and unlensed events using the likelihood and its corresponding odds ratio similar to \citep{Christian2018}, which works as a threshold of when a distinctive lensed model is achievable with a $FF= 0.97$. And we find that it returns similar thresholds to the $FF$, i.e. 
\begin{equation}
    y_c\approx 0.1 - 0.3~\rm{and}~M_{\rm L,z}\sim M_{\rm BBH, c}(1+z_{\rm S, c}),\label{eq:Oddsthreshold}
\end{equation} 
This conforms with the $FF$ analyses (Eq. \ref{eq:FFthreshold}). 
In Fig.~\ref{fig_odds}, we show an example of our odds ratio analysis with $y=0.1$. The GW events are assumed to be located at the horizon redshift to make $\rho(M_{\rm BBH,z}=42.5M_{\odot})=8.0$, so that the high-mass events are of SNRs $\rho>8$. 
The behavior of the odds ratio is quite similar in each case, showing that the lensing effect is easier to identify for more massive lenses. And overall, the analysis indicates that the lensing effect by PBHs of $M_{\rm L,z}\gtrsim M_{\rm BBH, c}(1+z_{\rm S, c})$ can be distinguished for the GW events of $M_{\rm BBH}\geq M_{\rm BBH, c}$ at $z_{\rm S, c}$. 

To summarize, for a given redshift $z_{\rm S, c}$, we can calculate the critical mass $M_{\rm BBH, c}$ by solving $\rho(M_{\rm BBH, c},z_{\rm S, c})=8$. While the $FF$ and odds ratio analyses show that all the detectable GW events within $z_{\rm S, c}$ (i.e. those of $M_{\rm BBH}\geq M_{\rm BBH, c}$) can be used to probe the possible lensing effect by point-mass lenses if their redshifted mass satisfy $M_{\rm L,z}\geq M_{\rm BBH, c}(1+z_{\rm S, c})$.

\section{The lensing probability}\label{probablity}

We now consider the probability of observing a lensed GW event.
We adopt the merger rate density with the form $R(z)=R_0(1+z)^\lambda$, while $R_0=52.9^{+55.6}_{-27.0}$ Gpc$^{-3}$ yr$^{-1}$, and for the quite uncertain parameter $\lambda$, we adopt two values ($0.9^{+9.8}_{-10.8}$ and $6.5^{+9.1}_{-9.3}$), based on the results presented in GWTC-1 \citep{ligo2018}. 
We also study the case of $\lambda\sim3$, which relates to the observed star formation rate \citep{Madau2014}. 

We model the mass function of BBHs following \citet{ligo2018},% with the probability distribution function of the primary BH, 
\be\label{eq:m1pdf}
p(m_1,m_2)\propto m_1^{-\alpha} q^\beta,
\ee
where $\alpha=-1.6^{+1.5}_{-1.7}$ \citep{Kroupa2001,ligo2018}, $q=m_2/m_1$, $\beta=6.7^{+4.8}_{-5.9}$, $M_{\rm min}\leq m_2\leq m_1\leq M_{\rm max}$, 
$M_{\rm min}=5M_\odot$ is the minimum mass obtained from both theoretical predictions and observations 
\citep{Bailyn1998,Ozel2010,Farr2011,Belczynski2012,Fryer2012,Kochanek2014}, 
and the maximum mass is $42_{-5}^{+12} M_{\odot}$ \citep{Heger2002,Belczynski2016,ligo2018}\footnote{Although a BBH merger (GW190521) with a total mass of $\sim150M_\odot$ has been detected, the origin of such a BBH system is still unclear \citep[see more details in][]{GW190521origin}. Therefore, its effect on the BBH mass distribution is not included in this work.}.

The lensing optical depth for a GW source at redshift $z_{\rm S}$ is given by \citep{Schneider1992} 
\ba
  &\tau(f_{\rm PBH},z_{\rm S},M_{\rm L,c}) = \frac{3}{2} f_{\rm PBH}\Omega_{\rm DM}  
  \nonumber\\
  &\int^{z_{\rm S}}_{0} {\rm d}z_{\rm L} \int_0^{y_c} dy
  \frac{(1+z_{\rm L})^2}{H(z_{\rm L})/H_0} 
  \frac{H_0 D_A(z_{\rm L},z_{\rm S})~ D_A(z_{\rm L})}
  {c D_A(z_{\rm S})} 2y ,%\big(M_{\rm L}(1+z_{\rm L})\big)
\ea
where the dark matter density $\Omega_{\rm DM}$ and the Hubble constant $H_0$ are taken from the recent Planck observations \citep{Planck2016}. 
The Hubble parameter and angular diameter distance at redshift $z$ are $H(z)$ and $D_A(z)$. 
Note this expression assumes that PBHs are distributed uniformly in the Universe. 

The expected lensing event number by PBHs is then given by
\ba
&N_{\rm L}(f_{\rm PBH})=\int_{0}^{z_{\rm S, c}} {\rm d} z_{\rm S} \int_{m_{\rm 1,min}}^{M_{\rm max}} {\rm d}m_1 \int_{m_{\rm 2,min }}^{m_1} {\rm d} m_2~ 
\nonumber\\
&R(z_{\rm S}) {{\rm d} V_{\rm c}\over {\rm d} z_{\rm S}} p(m1,m2) \tau(f_{\rm PBH},z_{\rm S},M_{\rm L,c}){T_{\rm obs}\over 1+z_{\rm S}},
\ea
where $m_{\rm 1,min}=\max[M_{\rm min}, 0.5M_{\rm BBH, c}]$, $m_{\rm 2,min}= \max[M_{\rm min}, M_{\rm BBH, c}-m_1]$, $V_{\rm c}$ is the comoving volume, and $T_{\rm obs}$ is the observational time. 
The lensing probability is calculated with $(1-e^{-\tau})\approx\tau$. 
Since the BBH merger rate formed by stellar black holes is likely to decrease at the redshift $z>3$ \citep{Carl2018,Fragione2018}, we assume $R(z>3)=0$ for ET-D. Therefore, the actual rate for ET-D can be higher if a non-zero merger rate at $z>3$ is considered.

\section{Results}\label{constraint}

Using the formalism presented in the previous section, we now demonstrate how future GW observations would lead to a tight constraint on the ratio $f_{\rm PBH}$.
The expected lensing event number is $N_{\rm L}\propto f_{\rm PBH}$, if a {\em null detection\/} of lensing is obtained after a period of observational run, one can place a constraint from $N_{\rm L}(f_{\rm PBH})<1$.
The constraint is mostly sensitive to the redshifted lensed mass $M_{\rm PBH,z}=M_{\rm PBH}(1+z)$, and to obtain the direct constraint on the PBH, we concentrate on the effective mass $M_{\rm PBH}=M_{\rm PBH,z}/(1+z_{\rm eff})$, where the effective redshift is defined as 
\begin{equation}
    z_{\rm eff}= \int z_{\rm S} R(z_{\rm S})\tau(z_{\rm S}) {{\rm d} V_{\rm c}\over {\rm d} z_{\rm S}} /\int R(z_{\rm S})\tau(z_{\rm S}) {{\rm d} V_{\rm c}\over {\rm d} z_{\rm S}}.
\end{equation}
As shown in Eq. (\ref{eq:FFthreshold} and \ref{eq:Oddsthreshold}), the lensing effect can only be identified for $M_{\rm L, z}\gtrsim M_{\rm BBH, c} (1+z_{\rm S, c})$, while the BBH merger events are of mass $M_{\rm BBH}\geq2M_{\rm min}=10M_\odot$ in our model. Consequently, the PBHs that can be constrained would also have mass $M_{\rm PBH}\geq10M_\odot$.

\begin{figure}
	\centering
	\includegraphics[width=0.49\textwidth]{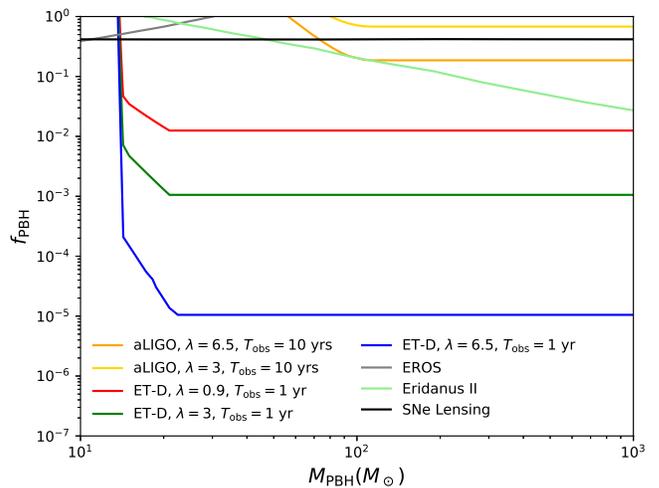}
	\caption{The constraint on the ratio $f_{\rm PBH}$ as a function $M_{\rm PBH}$ is shown for the aLIGO and ET-D. The threshold is adopted at $y_c=0.1$ and $M_{\rm L,z} \gtrsim 1.3M_{\rm BBH,z}$. 
	The regions below the corresponding curves indicate that $N_{\rm L}(f_{\rm PBH})\leq1$. 
	We also show the current constraint on $f_{\rm PBH}$ from the SNe lensing data \citep{Zum2018}, EROS \citep{Tisserand2007}, and Eridanus II \citep{Brandt2016,Li2017}.
	}\label{fig_pbh}
\end{figure}

The lensing events ($N_{\rm L}$) also depends on the impact parameter, i.e. $N_{\rm L}\propto y_c^2$. 
Adopting $y_c=0.1$ as an example, we show the constraints on $f_{\rm PBH}$ in Fig. \ref{fig_pbh} for different choices of rate parameter $\lambda$, where the observation time is assumed to be $T_{\rm obs}=1$ yr for the Einstein Telescope (ET), and $T_{\rm obs}=10$ yrs for the aLIGO. 
For the case of aLIGO, we obtain a $f_{\rm PBH}<0.2(y_c/0.1)^{-2}$ for $M_{\rm PBH}>50 M_\odot$. 
For the case of ET-D, on the other hand, we can expect from a null result a very tight constraint of $f_{\rm PBH}<(10^{-2}-10^{-5})(y_c/0.1)^{-2}$ for PBHs with mass $>10 M_{\odot}$. 

However, once a lensed GW signal is identified from observational data, the best-fitted values and errors can be obtained for the model parameters. We now examine the relative errors of lensing parameters from GW observation. We adopt the root-mean-square ($rms$) errors $\Delta \theta_p$ by inversing the Fisher Information Matrix (FIM, denoted by $\Gamma^{ij}$), 
%i.e. $\langle \Delta \theta_p^i \Delta \theta_p^j \rangle = \left( \Gamma^{-1} \right)^{ij}$.  
i.e. $\Delta \theta_p^i = \sqrt{\Sigma^{ii}}$ with $\Sigma^{i,j} = \left( \Gamma^{-1} \right)^{ij}$.
The FIM is given by \citep{Takahashi2003,Finn1992,Cutler1994},  
\be
\Gamma^{ij} = 4  {\mbox{Re}} \int \frac{{\rm d}f}{S(f)}~
\frac{\partial h_l^{ *}(f)}{\partial \theta_p^i}
\frac{\partial h_l(f)}{\partial \theta_p^j}.\label{eq:FIM}
\ee
For a wide range of parameters that we are interested in, we found $\Delta M_{\rm L,z}/M_{\rm L,z}\lesssim0.3$ and $\Delta y/y\lesssim13$. One can see that a good parameter estimation for the lens mass can be obtained. In the limit of geometrical optics, it is found that $\Delta M_{\rm L,z}/M_{\rm L,z}$ and $\Delta y/y$ are inversely proportional to the SNR \citep{Takahashi2003}, therefore better parameter estimation is expected if the lensed event is detected at higher SNR.

\section{conclusion and discussion}\label{conclusion}

In this paper, we have explored the possibility of detecting PBHs by searching for their microlensing effect on GW events. 
We assumed an uniform space distribution for the PBHs in the Universe, and fixed $\eta=0.238$ and $\chi=0.024$ for the BBH merger events following the GWTC-1 \citep{Ligo2018a} and GWTC-2 catalogs \citep{Abbott20202nd}.
The criteria, $FF= 0.97$ and $\log \rm{Odds}= 4$, are used to indicate whether such lensing signals are distinguishable. 
We used the Fisher information matrix to assess the precision of parameter estimation, under the positive detection scenario for a lensed GW event, where the PBH mass can be obtained with a relative error $\Delta M_{\rm PBH}/M_{\rm PBH}\lesssim0.3$.

In the case of a null detection of such events over a long observational time, we obtained tight constraints on $f_{\rm PBH}$ -- especially in the third-generation GW detector era. 
We found that, after one year of running, the ET-D can provide a very tight constraint of $f_{\rm PBH}\lesssim (10^{-2}-10^{-5})(y_c/0.1)^{-2}$ for $M_{\rm PBH}\gtrsim 10 M_\odot$. 
For aLIGO, on the other hand, the constraint is $f_{\rm PBH}\lesssim 0.2(y_c/0.1)^{-2}$ for PBHs with a mass greater than $\sim 50 M_\odot$ after ten years of running, weaker than that in Ref. \citep{Jung2019}. This is because we use a more conservative and realistic choice for the distinguishability criterion, as discussed in Section \ref{sec:intro}. 
The $FF$ and odds-ratio analyses show that $y_c\approx 0.1-0.3$. 
It can be seen that if one adopts a lensing parameter of $y_c=0.1$, a significant improvement can be obtained upon the recently updated constraints from the lensing of type Ia supernovae, namely, $f_{\rm PBH}\lesssim 0.35$ \citep{Zum2018}. 
And this constraint can be even tighter if a significant fraction of the GW events are detected with higher SNR.

It should be noted that we ignored the macrolensing effect from the structures hosting PBHs, such as galaxies and dark-matter halos. This treatment is appropriate for the low-mass dark-matter halos, such as those with mass $(10^6-10^8) M_\odot$, but in the presence of strong macrolensing by hosting galaxies, the microlensing effect can be significantly modified. In this case, the caustic shape of a point-mass lens would change from a single point to a diamond shape, leading to the formation of more images \citep[e.g.][]{Diego2019,Diego2020}, and thus Eq.~(\ref{eq:F_f}) needs to be modified. 
Recently, Cheung et al. 2021 \citep{Cheung2021} studied this effect in details and considered the microlensing effect in the presence of strong macrolensing, where the microlens is placed within the Einstein radius of the macrolens. In this case, the microlensing amplification factor can change significantly from that of an isolated microlens in the high-frequency ($>200$ Hz) band, as shown in their Fig. 4. 
And thus the $FF$ behavior is slightly different from our result. In their Fig. 9, the mismatch, which can be treated as $1-FF$, is found to be of variation at some lens mass, and impact parameter. As from our result, the mismatch will be $1-FF\gtrsim0.03$ when Eq. (\ref{eq:FFthreshold}) is satisfied. With $z_{\rm S}=1$, $M_{\rm BBH}=60M_\odot$, and $z_{\rm L}=0.5$ in \citep{Cheung2021}, this requires to $M_{\rm L} \gtrsim (80-128) M_\odot$ and $y<0.1-0.3$, which is also consistent with the high mismatch part of their Fig. 9.

Besides, stars in hosting galaxies could induce stellar microlensing. For strong macrolensing by Milky-Way-like galaxies, its probability may approach $P_{\rm Micro}\sim10^{-2}P_{\rm Macro}$ \citep[e.g.][]{Christian2018}, where $P_{\rm Macro}$ is the macrolensing probability. For a typical strong macrolensing probability of $P_{\rm Macro}\sim10^{-4}$, the associated microlensing a probability of $P_{\rm Micro}\sim10^{-6}$. While for more massive hosting galaxies, this rate may be enhanced significantly \citep[e.g.][]{Christian2018}. As such, if the stellar lensing effect cannot be distinguished from the point-mass lensing effect from the data analysis, the constraint on PBH abundance may be contaminated. 
Consequently, a more accurate treatment involving all these effects is required to improve the constraint on PBHs. This will be examined more in further work.

Nevertheless, by collaborating with space-borne GW detectors, such as TianQin \citep{Luo2016,Hu2017,Hu2019}, laser interferometer space antenna (LISA) \citep{Amaro2017}, and/or deci-hertz interferometer GW observatory (DECIGO) \citep{Sato2017}, it will become possible to study such lensed events through multi-band GW observation. 
This approach will have the potential to significantly further improve the constraint on $f_{\rm PBH}$. Space-borne GW detectors are expected to observe the early inspiral signals of BBH systems months to years before their final merger, thus allowing the determination of their intrinsic parameters with very high precision. Since the lensing affects the GW waveform differently at different frequencies, lensing will result in an obvious inconsistency between the best-fit parameters derived from space-borne and ground-based facilities -- thus providing a clear indicator for the existence of lensing.

%%%%%%%%%%%%%%%%%%%%%%%%%%%%%%%%%%%%%%%%%%%%%
%%%%%%%%%%%%%%%%%%%%%%%%%%%%%%%%%%%%%%%%%%%%%

\begin{acknowledgments}
We thank the referee for the useful suggestions and comments.
We thank Prof. M.A. Hendry, Prof. P.J. Zhang, Prof. D. Lai, Dr. J.X. Han, and Dr. Y.F. Wang for helpful discussions. 
J.S.W. was supported by China Postdoctoral Science Foundation. Y.M.H. is supported by the National Key Research and Development Program of China (No. 2020YFC2201400) and Guangdong Major Project of Basic and Applied Basic Research (Grant No. 2019B030302001).
\end{acknowledgments}

\appendix

\section{Validity of the {\it FF=0.97} threshold}\label{appendix:FF}

If the lensing effect is ignored in the matched filtering analysis of GW signals, the best-matched waveform $h_s(f,\theta_p+\delta \theta_p)$ would be shifted from the true parameter $h(f,\theta_p)F(f,\theta_l)$ by an amount of $\delta \theta_p$. 
Meanwhile, the parameters estimated from the data analyses are accompanied by uncertainties of $\Delta \theta_p$, which are calculated with the Fisher Information Matrix (Eq. \ref{eq:FIM}) here.
%The comparison between the shift and uncertainty serves as an additional channel to identify the existence of the lens PBH. 
We consider a case with $\rho=8$ for the original unlensed GW waveform with $M_{\rm BBH,z}=42.5M_\odot$ and use the PSD of aLIGO. 
Then we study how much difference is induced by a lensing object with $y=0.1$ and $M_{\rm L,z} \gtrsim44 M_\odot$ (corresponding to $FF\lesssim0.97$). 
We found that the best-matched waveform (i.e. the `shifted' waveform) corresponds to $\rho_s\simeq8$, effectively no obvious difference. 
This can also be seen in Fig. \ref{fig_lens}, where the best-matched waveforms mimic the original unlensed waveforms. 
But we found that the lensing effect could alter these intrinsic parameters in some cases, especially for the symmetry mass ratio, whose shifted values would be larger than the $rms$ error, i.e. $\delta \eta\gtrsim\Delta \eta$.
However, if we take into consideration the lensing effect, we found the lensed waveform would be of $\rho_l\gtrsim 11$, which means the failure of including the lensing effect would greatly decrease the SNR.

Also, we can study the AIC \citep{Akaike1974} for the `shifted' waveform model and the lensed waveform model. 
A difference of 6 in the AIC is generally considered as strong evidence in favor of the model with a lower AIC \citep{Kass1995}. 
The AIC depends on the likelihood function $\ln\Lambda(h_1)=(h_o|h_1)-0.5(h_1|h_1)$ \citep{Jaranowski2012} of the waveform $h_1$, 
\be
{\rm AIC}(h_1)=-2\ln\Lambda(h_1)+2k=-2(h_o|h_1)+(h_1|h_1)+2k,
\ee
where we assume that a lensed GW signal is received by the detector, i.e. $h_o=h_l$ is the signal, and $k$ is the number of the parameters in the model. For the lensed waveform, we have ${\rm AIC}(h_l)=-\rho_l^2+2k_l$; 
while for the `shifted' waveform, we obtain ${\rm AIC}(h_s) =-2FF*\rho_l\rho_s+\rho_s^2+2k_s$. 
The difference is then 
\be
\Delta {\rm AIC}={\rm AIC}(h_l)-{\rm AIC}(h_s)=-\rho_l^2-\rho_s^2+2FF*\rho_l\rho_s+4,
\ee
where the lensed waveform has two more parameters $(M_{\rm L,z}, y)$, thereby $k_l-k_s=2$. 
Taking $y=0.1$ and $M_{\rm L,z}\gtrsim M_{\rm BBH,z}$ for different values of $M_{\rm BBH,z}$, we found this difference is generally $\Delta {\rm AIC}\leq -6.9$ for the lensed events with $FF\leq0.97$. 
This also confirms that the lensed waveform can be distinguished by real data analysis. 
%Very recently, it is found that aLIGO can detect such lensing event with $M_{\rm BBH,z}=65M_\odot$ and $M_{\rm L}\gtrsim 30M_\odot$ ($\gtrsim 60M_\odot$) when the SNR is larger than $\sim30~(\sim10)$ using a thorough Bayesian model selection pipeline.\citep{Christian2018}. A $FF\approx0.97$ with $y=0.1$ and $M_{\rm L}\approx60M_\odot$ is also concluded from this study, consistent with their finding.

%\nocite{*}
\bibliography{ref}
\end{document}